\documentclass[12pt]{article}
\usepackage{amssymb,amsmath,amsthm}
\newcommand{\Real}{\mathbb{R}}
\newcommand{\pd}{\partial}
\newcommand{\D}{\mathrm{d}}
\newcommand{\Nat}{\mathbb{N}}

\newcommand{\Hilbert}{\mathcal{H}}

\newcommand{\supp}{\mathrm{supp}}

\newcommand{\eps}{\varepsilon}

\newtheorem{Lemma}{Lemma}
\newtheorem{Theorem}{Theorem}

\theoremstyle{definition}

\newtheorem{ass}{Assumption}

\begin{document}
%
\title{\textbf{\Large
Spectrum of the Schr\"odinger operator in \\ a perturbed
periodically twisted tube }}
\author{
P.~Exner$^{a,b}$ and H.~Kova\v{r}{\'\i}k$^{a,c}$ }
\date{}
\maketitle
\begin{flushleft}
{\small \em a) Department of Theoretical Physics, Nuclear Physics
Institute, Czech Academy \\ $\phantom{a)}$ of Sciences, CZ-25068
\v{R}e\v{z} near Prague \\ b) Doppler Institute, Czech Technical
University, B\v{r}ehov{\'a} 7, CZ-11519 Prague \\
c) Institute for Analysis, Dynamics and Modeling, Faculty of Mathematics and \\
$\phantom{a)}$ Physics, Stuttgart University, PF 80 11 40, D-70569 Stuttgart. Germany \\
 $\phantom{a)}$ {\rm exner@ujf.cas.cz, kovarik@mathematik.uni-stuttgart.de}}
\end{flushleft}
%
\begin{abstract}
\noindent We study Dirichlet Laplacian in a screw-shaped region,
i.e. a straight twisted tube of a non-circular cross section. It
is shown that a local perturbation which consists of ``slowing
down'' the twisting in the mean gives rise to a non-empty discrete
spectrum.
\end{abstract}
%
%
%

\section{Introduction}\label{Sec.Intro}
%
Existence of geometrically-induced bound states in infinitely
extended regions of tubular shape was noticed at the end of the
eighties, first in the two-dimensional situation \cite{ES}, and
studied intensively since then -- see \cite{DE, GJ, LCM, RB}, and
more recently \cite{ChDFK, EFK}. The effective interaction which
lead in these cases to the existence of a discrete spectrum came
from tube bends. If the tube is embedded in $\Real^d$ with $d\ge
3$ one can ask about the effect of its twisting on the spectrum.
In the work quoted above the tubes considered were either circular
or twisted in a particular way aligned with the tube axis torsion;
in that case the twist did not affect the result in the leading order.

Investigations going beyond this special case were done only
recently. Inspired by the existence of a magnetic Hardy-type
inequality in waveguides \cite{EK} the authors of \cite{EKK}
analyzed the generic case of a local tube twist and found that it
gives rise to a repulsive effective interaction which can destroy
weakly bound states coming from other perturbations. In this
letter we push this idea further and study a twist which extends
along the whole tube.

As a repulsive interaction it rises the essential spectrum
threshold. At the same time, if we perturb such a screw-shaped
tube locally in such a way that the repulsion is weakened, e.g. by
a local slowdown of the twist, one may expect a binding effect. We
show, under mild regularity assumptions, that it is indeed the
case and that it is sufficient if the twisting is slowed down
locally in the mean. Moreover, a natural analogy with
one-dimensional Schr\"odinger operator theory suggests that the
effect might survive in the critical case when the mean value of
the twist variation is zero. It is not a reliable guide, of
course, because sometimes in similar situations critical bound
states are absent \cite{BEGK}, nevertheless, here we are able to
demonstrate that discrete spectrum for a critical twist
perturbation is still non-empty.

Let us review the contents of the letter. We will introduce the
needed notation in the next section, then we will analyze the
spectrum in the periodic case. Our main results are given in
Section~\ref{s: main}, specifically in Theorems~\ref{main} and
\ref{critical} for the non-critical and critical situation,
respectively.

\section{Preliminaries}

First we fix the notation. Let $\omega$ be an open bounded and
connected set in $\Real^2$ and let $\theta$ be a differentiable
function from $\Real$ to $\Real$. For $s\in\Real$ and
$t:=(t_2,t_3)\in\omega$ we define the mapping $\mathcal{L}$ from
$\Real\times\omega$ to $\Real^3$ by
\begin{equation} \label{mappingL}
\mathcal{L}(s,t) = (s,\, t_2\cos\theta(s)+t_3\sin\theta(s),\,
t_3\cos\theta(s)-t_2\sin\theta(s))\,.
\end{equation}
The image $\mathcal{L}(\Real\times\omega)$ is a tube in $\Real^3$
which is twisted unless the function $\theta$ is constant. A case
of particular interest is a screw-shaped tube corresponding to a
linear $\theta$. We fix a positive constant $\beta_0$ and define
the tube $\Omega_0$ by
 $$
 \Omega_0:= \mathcal{L}_0(\Real\times\omega)\,,
 $$
where
$$
 \mathcal{L}_0(s,t) :=  (s,\, t_2\cos(\beta_0 s)
 +t_3\sin(\beta_0 s),\,
 t_3\cos(\beta_0 s)-t_2\sin(\beta_0 s))\,;
 $$
it will play role of the unperturbed system. The operator we will be
concerned with is the Dirichlet Laplacian $H_0$ on
$L^2(\Omega_0)$, i.e. the self-adjoint operator associated with
the closed quadratic form
 \begin{equation} \label{qadfrom}
 Q_0[\psi] := \int_{\Omega_0}\, |\nabla\psi|^2\, \D s\,\D t\, ,
 \qquad \forall\, \psi\in
 D(Q_0)= \Hilbert^1_0(\Omega_0)\,.
 \end{equation}

\section{Spectrum of $H_0$}
Given $\psi\in C_0^{\infty}(\Real\times\omega)$ it is useful to
introduce the following shorthand,
\begin{equation} \label{twisting}
\psi'_{\tau} := t_2\pd_{t_3}\psi-t_3\pd_{t_2}\psi\, .
\end{equation}
A simple substitution of variables shows that
$$
Q_0[\psi]= \int_{\Real\times\omega}\,
|\nabla_t\psi|^2+|\pd_s\psi+\beta_0\psi'_{\tau}|^2\, \D s\,\D t\,
,
$$
where
$$
\nabla_t\psi:=(\pd_{t_2}\psi,\pd_{t_3}\psi)\, .
$$
In other words, the operator $H_0$ acts on its domain in $L^2(\Omega_0)$ as
$$
H_0= -\pd_{t_2}^2-\pd_{t_3}^2+(-i\pd_s- i\, \beta_0\,
(t_2\pd_{t_3}-t_3\pd_{t_2}))^2
$$
Since $\beta_0$ is independent of $s$ we are able to employ a
partial Fourier transformation $\mathcal{F}_s$ given by
$$
(\mathcal{F}_s\, {\psi})(p,t) = \hat{\psi}(p,t)=
\frac{1}{\sqrt{2\pi}}\, \int_{\Real}\, e^{-i\, ps}\, \psi(s,t)\D s
\,,
$$
which allows us ro rewrite the quadratic form as
$$
Q_0[\hat{\psi}] = \int_{\Real\times\omega}\,
|\nabla_t\hat{\psi}|^2+|i\, p\,
\hat{\psi}+\beta_0\hat{\psi}'_{\tau}|^2\, \D p\, \D t
$$
for a suitably regular $\psi$. Since the transformation
$\mathcal{F}_s$ extends to a unitary operator on
$L^2(\Real\times\omega)$, the operator $H_0$ is unitarily
equivalent to the direct integral
\begin{equation} \label{direct}
\int^{\oplus}_\Real\, h(p)\, \D p
\end{equation}
with the fibre operator
\begin{equation} \label{fibre}
h(p)= -\pd^2_{t_2}-\pd^2_{t_3}+(p-i\, \beta_0(t_2\pd_{t_3}\,
-t_3\pd_{t_2}))^2\,
\end{equation}
on $L^2({\omega})$ subject to Dirichlet boundary conditions at
$\partial\omega$. Introducing the polar coordinates $(r,\alpha)$
on $\omega$, we can rewrite $h(p)$ as follows
\begin{equation} \label{fiber}
h(p) = -\Delta_D^{\omega}+(p-i\, \beta_0\pd_{\alpha})^2\, ,
\end{equation}
where $-\Delta_D^{\omega}$ denotes the Dirichlet Laplacian in $r$
and $\alpha$. Since $h(p)$ is a sum of $-\Delta_D^{\omega}$ and a
positive perturbation, it follows easily from the minimax
principle that its spectrum is purely discrete. Let us denote the
eigenvalues of $h(p)$ by $\eps_n(p)$ and the respective
eigenfunctions by $\psi_n(p)$, i.e.
$$
h(p)\, \psi_n(p)=\eps_n(p)\, \psi_n(p)\, .
$$

\begin{Lemma} \label{bandfunctions}
Every $\eps_n(\cdot),\, n\in\Nat,$ is a real-analytic function of
$p$ and
\begin{equation} \label{limit}
\lim_{p\to\pm\infty}\, \eps_n(p) \to \infty\, .
\end{equation}
\end{Lemma}
\begin{proof}

It is not difficult to check that the quadratic form associated
with the operator $h(0)$ defined on the form domain
$\Hilbert^1_0(\Omega_0)$ is non-negative and closed.
This implies that $h(0)$ is self-adjoint on its
natural domain which we denote as $D(0)$. Let us formally expand
the square in (\ref{fiber}) and write $h(p)$ as
$$
h(p) = h(0) +p^2 -2i\, p\, \beta_0\,\pd_{\alpha} \, .
$$
Denote the resolvent of $h(0)$ at a point $z\in\mathbb{C}$ by
$R_z$, i.e. $R_z=(h(0)-z)^{-1}$. Then we have for any $\varphi\in
C_0^{\infty}(\omega)$ the following estimate
\begin{eqnarray*}
\|\pd_{\alpha}\varphi\|^2 & \!\leq\! & (\varphi,\, h(0)\varphi) =
(R_z(h(0)-z)\varphi,\, h(0)\varphi) \\
& \!\leq\! & \|R_z\|\, \|h(0)\varphi\|^2+|z|\, (\varphi,\,
R_{\bar{z}}\, h(0)\varphi) \\
& \!\leq\! & C(z)\, \|h(0)\varphi\|^2+|z|^2\, C(z)\,
\|\varphi\|^2\, ,
\end{eqnarray*}
where $C(z)\to 0$ as $\Im z\to \infty$. Consequently, $i\,
\pd_{\alpha}$ is $h(0)$-bounded with the relative bound zero which
implies that the domain of $h(p)$ coincides with $D(0)$ and the
vector $h(p)\phi$ is analytic as a function of $p$ for every
$\phi\in D(0)$ (since $p^2$ is clearly analytic). From
\cite{Kato}, pp. 375 and 385, it thus follows that $\{h(p):\:
p\in\Real\}$ is a self-adjoint analytic family of type $A$ and
that all the $\eps_n(\cdot)$ are real-analytic functions of $p$.

\noindent To prove the second statement of the lemma, let us first
define the cross-section radius with respect to the rotation axis,
$$
 a:= \sup_{t\in\omega}\, |t|\, .
$$
We observe that for any  $\varphi\in C_0^{\infty}(\omega)$ we have
a trivial pointwise inequality,
$$
|2p\, \beta_0\, \bar{\varphi}\, \pd_{\alpha}\varphi| \leq p^2\,
\frac{\beta_0^2}{\beta_0^2+a^{-2}}\,
|\varphi|^2+(\beta_0^2+a^{-2})\,
  |\pd_{\alpha}\varphi|^2\, ,
$$
which implies that
\begin{eqnarray*}
\lefteqn{(\varphi,\, h(p)\, \varphi) = \int_{\omega}\,
\left(|\pd_r\varphi|^2+\frac{1}{r^2}\,
|\pd_{\alpha}\varphi|^2+|(p-i\,
  \beta_0\, \pd_{\alpha})\varphi|^2\right )\, r\, \D r\, \D\alpha} \\
&& \geq \int_{\omega}\, \left(|\pd_r\varphi|^2+a^{-2}\,
  |\pd_{\alpha}\varphi|^2+p^2\, |\varphi|^2 - |2p\, \beta_0\, \bar{\varphi}\,
  \pd_{\alpha}\varphi|+\beta_0^2\, |\pd_{\alpha}\varphi|^2\right)
  r\, \D r\, \D\alpha \\
 && \geq  \frac{1}{1+a^2\, \beta_0^2}\, \, p^2\, \int_{\omega}\,
 |\varphi|^2\, r\, \D r\, \D\alpha\,;
\end{eqnarray*}
this in turn yields the sought result. \end{proof}

\noindent It is clear from (\ref{fibre}) that the spectral
threshold of $h(0)$ cannot be lower than that of
$-\Delta_D^{\omega}$. It has been shown in \cite{EKK} that the
inequality is sharp,
\begin{equation} \label{infimum}
E := \inf\sigma(h(0)) > \inf\sigma\left(-\Delta_D^{\omega}\right )\, ,
\end{equation}
whenever $\omega$ is \emph{not} rotationally symmetric. This
follows, by the way, also from our Lemma \ref{positivity}$b$ which
will be proved below. \par Our aim is to show that this quantity
determines the spectral threshold of our original Hamiltonian, in
other words, $E=\inf\sigma(H_0)$. To this end let us denote by $f$
the real-valued eigenfunction of $h(0)$ associated with the
eigenvalue $E=\eps_1(0)$, i.e.
\begin{equation} \label{grstate}
 h(0)f = -\Delta_D^{\omega}\, f-\beta_0^2\, \pd^2_{\alpha}f = E f\, .
\end{equation}
Then we can make the following claim.

\begin{Lemma} \label{positivity}
Let $f$ be given by (\ref{grstate}). Then
\begin{itemize}
\item[(a)] $f$ is strictly positive in $\omega$. \item[(b)]
$\int_{\omega}\, |f'_{\tau}|^2\, \D t = \int_{\omega}\,
|\pd_{\alpha}f|^2 \, \D t > 0$ provided  $\omega$ is not
rotationally symmetric.
\end{itemize}
\end{Lemma}

\begin{proof}
To prove the positivity of $f$ it is enough to show that the
semigroup $e^{-t\, h(0)}$ is positivity improving for all $t>0$,
see \cite[Thm~XIII.44]{RS4}, that is, we have to show that
$e^{-t\, h(0)}$ maps every positive function in $\omega$ into a
strictly positive function in $\omega$. Since $-\Delta_D^{\omega}$
commutes with $\pd^2_{\alpha}$, we get
$$
e^{-t\, h(0)} = e^{t\, \Delta_D^{\omega}}\, e^{t\, \beta_0^2\,
  \pd^2_{\alpha}}\, .
$$
However, it follows easily from \cite[Thm~XIII.50]{RS4} that
$e^{t\, \beta_0^2\, \pd^2_{\alpha}}$ is positivity preserving for
all $t>0$, i.e. it maps every positive function into a positive
function. Now note that since $-\Delta_D^{\omega}$ has a strictly
positive ground state, $e^{t\, \Delta_D^{\omega}}$ is positivity
improving for all $t>0$ by \cite[Thm~XIII.44]{RS4}. Hence given a
positive function $g$ in $\omega$, we know that $e^{t\,
\beta_0^2\, \pd^2_{\alpha}}g$ is positive, which means that
$e^{-t\, h(0)}g$ is strictly positive; this proves the first
statement of the Lemma.

The second statement is an immediate consequence of the first one.
Let $B$ be the biggest circle (i.e., the one with the biggest
radius) centred at the origin, such that $B\subset
\overline{\omega}$. Denote its complement in $\overline{\omega}$
by $B^c$. By assumption we know that $B^c\neq\emptyset$. Since $f$
satisfies Dirichlet boundary conditions on $\partial\omega$ and is
strictly positive inside $\omega$, it follows that
$|\partial_{\alpha}f|$ is strictly positive in almost every point
of $B^c\cap\partial\omega$, where $\partial\omega$ is not a part
of a circle centred at the origin. This ''non-circular'' part is,
of course, a positive measure set, hence using the
differentiability of $f$ we can find a neighbourhood of
$B^c\cap\partial\omega$ with a positive Lebesgue measure on which
$|\partial_{\alpha}f|>0$.
\end{proof}

\noindent {\bf Remark:} The first statement of Lemma \ref{positivity} also
follows from \cite[Thm.~8.38]{GT}.

\vspace{0.2cm}

\noindent Now we are able to determine the spectrum of the free operator.

\begin{Theorem}
The spectrum of $H_0$ is purely absolutely continuous and covers
the half-line $[E,\infty)$, where $E$ is the lowest eigenvalue of
$h(0)$.
\end{Theorem}

\begin{proof}
From (\ref{direct}) and Lemma \ref{bandfunctions} we know that the spectrum of
$H_0$ is absolutely continuous and that $[E,\infty) \subset\sigma(H_0)$. It
remains to show that
\begin{equation} \label{positive}
(-\infty,\, E) \cap \sigma(H_0) =\emptyset\, .
\end{equation}
Using the fact that the ground-state eigenfunction $f$ is strictly
positive in $\omega$, we can decompose any $\psi\in
C_0^{\infty}(\omega)$ as
\begin{equation} \label{decomposition}
\psi(s,t) = f(t)\varphi(s,t)\, .
\end{equation}
We use the fact $f$ is real-valued  and integrate by parts to get
\begin{eqnarray*}
Q_0[\psi]-E\, \|\psi\|^2 & = & \int_{\Real\times\omega}\, \Big(
f^2\, |\nabla_t\varphi|^2
  -(\Delta_D^{\omega}f)f|\varphi|^2+f^2\, |\pd_s\varphi|^2 \\
&& +\beta_0\, f\pd_{\alpha}f(\pd_s\bar{\varphi}\,
\varphi+\bar{\varphi}\, \pd_s\varphi) +\beta_0\,
f^2(\pd_s\bar{\varphi}\,
\pd_{\alpha}\varphi+\pd_{\alpha}\bar{\varphi}\,
\pd_s\varphi) \\
&& +\beta_0^2\, f^2\, |\pd_{\alpha}\varphi|^2-\beta_0^2\,
(\pd_{\alpha}^2f)f|\varphi|^2 -E\, f^2\, |\varphi|^2 \Big)\, \D
s\, \D t \, .
\end{eqnarray*}
Since
$$
\int_{\Real}\, (\pd_s\bar{\varphi}\,
\varphi+\bar{\varphi}\, \pd_s\varphi)\, \D s = 0
$$
and
$$
-\Delta_D^{\omega} f-\beta_0^2\, \pd_{\alpha}^2 f -E\, f = 0,
$$
see (\ref{grstate}), we finally obtain
$$
Q_0[\psi]-E\, \|\psi\|^2 = \int_{\Real\times\omega}\,
f^2\left(|\nabla_t\varphi|^2+ |\pd_s\varphi+\beta_0\,
\varphi'_{\tau}|^2\right )\, \D s\, \D t\, \geq 0\, .
$$
This implies (\ref{positive}).
\end{proof}

\section{Local perturbations of the twisting} \label{s: main}

After analyzing the ``free'' case, where the twisting velocity
$\dot\theta$ was constant, we want to look now what will happen if
the translation invariance of our tube is broken. We will suppose
that the velocity of the twisting is given by
\begin{equation} \label{twist}
\dot\theta(s) = \beta_0 - \beta(s)\, ,
\end{equation}
where $\beta(\cdot)$ is a bounded function such that $\supp\,
\beta \subset [-s_0,s_0]$ for some $s_0>0$. Let $\Omega_{\beta}$
denote the corresponding tube being defined by
$$
\Omega_{\beta}:= \mathcal{L}(\Real\times\omega)\,,
$$
where $\mathcal{L}$ refers to the twisting obtained by integration
of (\ref{twist}). We use the symbol $H_\beta$ for the Dirichlet
Laplacian on $L^2(\Omega_{\beta})$ and
\begin{equation} \label{form2}
Q_{\beta}[\psi] := \int_{\Omega_{\beta}}\, |\nabla\psi|^2\, ,
\end{equation}
will be the associated quadratic form with the form domain
$D(Q_{\beta})=\Hilbert_0^1(\Omega_{\beta})$. Since the support of
the perturbation $\beta(s)$ is compact, it is straightforward to
check that
\begin{equation} \label{essential}
\sigma_{ess}(H_\beta) = \sigma_{ess}(H_0) =[E,\, \infty) \, .
\end{equation}
Our main result says that if the tube twisting is locally slowed
down in the mean, the discrete spectrum of $H_\beta$ is non-empty.

\begin{Theorem} \label{main}
Assume that $\omega$ is not rotationally symmetric and that
\begin{equation} \label{ass}
\int_{-s_0}^{s_0}\, (\dot\theta^2(s)-\beta_0^2)\, \D s <0\, ,
\end{equation}
where $\dot\theta(\cdot)$ is given by (\ref{twist}). Then the
operator $H_\beta$ has at least one eigenvalue of finite
multiplicity below the threshold of the essential spectrum.
\end{Theorem}
\begin{proof}
Following the idea of \cite{GJ} we start constructing a trial
function from a transverse eigenfunction corresponding to the
bottom of the essential spectrum. Given $\delta>0$ we put
$\Psi_{\delta}(s,t)=f(t)\, \varphi(s)$, where
\begin{equation}
\varphi(s) = \left\{
\begin{array}{l@{\quad \mathrm{if} \quad }l}
 e^{\delta\, (s_0+s)} & s \leq -s_0\, , \\
 1 & -s_0\leq s \leq  s_0\, , \\
 e^{-\delta\, (s-s_0)} & s \geq s_0\, .
\end{array}
\right.
\end{equation}
It is easy to see that $\Psi_{\delta}\in D(Q_{\beta})$. A
straightforward calculation then gives
$$
Q_{\beta}[\Psi_{\delta}]-E\, \|\Psi_{\delta}\|^2 = \delta\,
\|f\|^2_{L^2(\omega)}-\|f'_{\tau}\|^2_{L^2(\omega)}\,
\int_{-s_0}^{s_0}\, (\dot\theta^2(s)-\beta_0^2)\, \D s
$$
and
$$
\|\Psi_{\delta}\|^2 = (\delta^{-1}+2s_0)\, \|f\|^2_{L^2(\omega)}\, .
$$
For $\delta\to 0$ we then get
$$
\frac{Q_{\beta}[\Psi_{\delta}]-E\,
\|\Psi_{\delta}\|^2}{\|\Psi_{\delta}\|^2}\, =\, \delta\,
\frac{\|f'_{\tau}\|^2_{L^2(\omega)}}{\|f\|^2_{L^2(\omega)}}\,
\int_{-s_0}^{s_0}\, (\dot\theta^2(s)-\beta_0^2)\, \D s +
\mathcal{O}(\delta^2)\, .
$$
Thus in view of Lemma \ref{positivity}$b$ it is sufficient to
choose $\delta$ small enough to achieve
$$
\frac{Q_{\beta}[\Psi_{\delta}]-E\,
\|\Psi_{\delta}\|^2}{\|\Psi_{\delta}\|^2}\, <0
$$
and the claim of the theorem follows.
\end{proof}

Validity of the above result can be extended also to the critical
case when the integral in (\ref{ass}) vanishes, however, we need a
somewhat stronger assumption on the regularity of $\dot\theta$. We
also have to suppose that the twisting is ``not fully reverted''
by the perturbation.

\begin{Theorem} \label{critical}
Assume that $\omega$ is not rotationally symmetric and let
$\dot\theta(\cdot)$ be given by (\ref{twist}). Suppose in addition
that $\dot\theta(s)+\beta_0>0$ holds for $|s|\le s_0$, and that
$\ddot\theta$ exists and is of the class $L^2([-s_0,s_0])$. Let
\begin{equation} \label{ass2}
\int_{-s_0}^{s_0}\, (\dot\theta^2(s)-\beta_0^2)\, \D s =0\, ;
\end{equation}
then the operator $H_\beta$ has at least one eigenvalue of finite
multiplicity below the threshold of the essential spectrum.
\end{Theorem}

\begin{proof}
Following again the idea of \cite{GJ} we improve the trial
function used in the proof of Theorem \ref{main} by a deformation
in the central region,
$$
\Psi_{\delta, \gamma}(s,t):=f(t)\, \varphi_{\gamma}(s)\, ,
$$
where
\begin{equation}
\varphi_{\gamma}(s) = \left\{
\begin{array}{l@{\quad \mathrm{if} \quad }l}
 e^{\delta\, (s_0+s)} & s \leq -s_0\, , \\
 1+\gamma\, (\beta_0-\dot\theta(s)) & -s_0\leq s \leq  s_0\, , \\
 e^{-\delta\, (s-s_0)} & s \geq s_0\, .
\end{array}
\right.
\end{equation}
with $\gamma>0$. Similarly as in the proof of Theorem \ref{main} one can check
that
$$
Q_{\beta}[\Psi_{\delta,\gamma}]-E\, \|\Psi_{\delta,\gamma}\|^2 =
\int_{\Real\times\omega}\, \left(\varphi_{\gamma}^2\,
  (f'_{\tau})^2\left(\dot\theta^2(s)-\beta_0^2\right)+f^2\,
  (\varphi'_{\gamma})^2\right)\,
\D s\, \D t\, .
$$
Using the assumptions of the theorem we find that the integrals
appearing in the last expression behave as
$$
\int_{-s_0}^{s_0}\, \varphi_{\gamma}^2\,
  \left(\dot\theta^2(s)-\beta_0^2\right)\, \D s =
  -2\gamma\int_{-s_0}^{s_0}\,
    \left(\dot\theta(s)-\beta_0\right)^2\left(\dot\theta(s)
    +\beta_0\right)\, \D s + \mathcal{O}(\gamma^2)\, ,
$$
and
\begin{eqnarray*}
\int_{\Real}\, (\varphi'_{\gamma})^2\, \D s & = & \delta+\gamma^2
\int_{-s_0}^{s_0}\, \left(\ddot\theta(s)\right)^2\, \D s
 = \mathcal{O}(\gamma^2)+\mathcal{O}(\delta)\, .
\end{eqnarray*}
as $\gamma,\, \delta\to 0$; the last two equations then give
 \begin{eqnarray*}
 \frac{Q_{\beta}[\Psi_{\delta,\gamma}]-E\,
  \|\Psi_{\delta,\gamma}\|^2}{\|\Psi_{\delta,\gamma}\|^2}
 & = & -2\, \gamma\, \delta\, \frac{\|f'_{\tau}\|^2_{L^2(\omega)}}{\|f\|^2_{L^2(\omega)}}\,
 \int_{-s_0}^{s_0}\, \left(\dot\theta(s)-\beta_0\right)^2
 \left(\dot\theta(s)+\beta_0\right)\, \D s  \\
 && + \delta\, \mathcal{O}(\gamma^2)+\mathcal{O}(\delta^2)\,.
 \end{eqnarray*}
Setting now $\gamma=\sqrt{\delta}$ we see that it is enough to
take $\delta$ small enough to get
$$
\frac{Q_{\beta}[\Psi_{\delta,\gamma}]-E\,
  \|\Psi_{\delta,\gamma}\|^2}{\|\Psi_{\delta,\gamma}\|^2} < 0\, ,
$$
which concludes the proof.
\end{proof}


\subsection*{Acknowledgments}

The research has been partially supported by Czech Academy od
Sciences and its Grant Agency within the projects IRP AV0Z10480505
and A100480501, and by DAAD within the project D-CZ~5/05-06.

%


\end{document}